\documentclass[10pt,twocolumn,english,aps,prl,showpacs,showkeys]{revtex4}
\usepackage[T1]{fontenc}
\usepackage[latin9]{inputenc}
\usepackage{babel}

\usepackage{prettyref}
\usepackage{amsthm}
\usepackage{amsmath}
\usepackage[unicode=true, 
 bookmarks=true,bookmarksnumbered=false,bookmarksopen=false,
 breaklinks=false,pdfborder={0 0 0},backref=false,colorlinks=false]
 {hyperref}
\hypersetup{pdftitle={Bures measure of entanglement of an arbitrary state of two qubits},
 pdfauthor={Alexander Streltsov}}
 
\makeatletter
\@ifundefined{textcolor}{}
{%
 \definecolor{BLACK}{gray}{0}
 \definecolor{WHITE}{gray}{1}
 \definecolor{RED}{rgb}{1,0,0}
 \definecolor{GREEN}{rgb}{0,1,0}
 \definecolor{BLUE}{rgb}{0,0,1}
 \definecolor{CYAN}{cmyk}{1,0,0,0}
 \definecolor{MAGENTA}{cmyk}{0,1,0,0}
 \definecolor{YELLOW}{cmyk}{0,0,1,0}
 }
  \theoremstyle{plain}
  \newtheorem*{thm*}{Theorem}

\usepackage{lmodern}
\newcommand{\bra}[1]{\langle #1|}
\newcommand{\ket}[1]{|#1\rangle}
\newcommand{\braket}[1]{\langle#1\rangle}

\makeatother

\begin{document}

\title{Bures measure of entanglement of an arbitrary state of two qubits}

\author{Alexander Streltsov}

\email{alexander.streltsov@physik.uni-wuerzburg.de}

\affiliation{Universit\"{a}t W\"{u}rzburg, Institut f\"{u}r Theoretische Physik und Astrophysik,
97074 W\"{u}rzburg, Germany}
\begin{abstract}
In {[}Phys. Rev. Lett. \textbf{80}, 2245 (1998){]} an explicit expression
for entanglement of formation for any two qubit state was given. Based
on this result we present an expression for the Bures measure of entanglement
for two qubit states. This measure was proposed in {[}Phys. Rev. A
\textbf{57}, 1619 (1998){]}, where the authors showed that it satisfies
all properties every entanglement measure must have.
\end{abstract}



\maketitle
A bipartite mixed state $\rho$ on a Hilbert space $\mathcal{H}=\mathcal{H}_{A}\otimes\mathcal{H}_{B}$
is in general called entangled if it can not be written in the form
\begin{equation}
\rho=\sum_{i}p_{i}\rho_{i}^{A}\otimes\rho_{i}^{B},\label{eq:rho}\end{equation}
with nonnegative probabilities $p_{i}$, $\sum_{i}p_{i}=1$, and $\rho_{i}^{A,B}$
being states on $\mathcal{H}_{A,B}$. Otherwise the state is called
separable.

Entanglement of pure bipartite states $\ket{\psi}$ is usually quantified
by the entanglement entropy \begin{eqnarray*}
E\left(\ket{\psi}\right) & = & -Tr\left[\rho^{A}\log_{2}\rho^{A}\right],\\
\rho^{A} & = & Tr_{B}\left[\ket{\psi}\bra{\psi}\right].\end{eqnarray*}
For mixed states many different measures were proposed, two of them
are \emph{entanglement of formation} $E_{F}$ and \emph{Bures measures
of entanglement} $E_{B}$ \cite{Horodecki2009}.

$E_{F}$ is defined as the minimal entanglement needed on average
to create the state: \[
E_{F}\left(\rho\right)=\min_{\rho=\sum_{i}p_{i}\ket{\psi_{i}}\bra{\psi_{i}}}\sum_{i}p_{i}E\left(\ket{\psi_{i}}\right).\]
For two qubit states a simple expression is known, it can be found
in \cite{Wootters1998}.

Bures measure of entanglement was proposed in \cite{Vedral1998} and
is defined as the Bures distance from the set of separable states:
\begin{equation}
E_{B}\left(\rho\right)=\min_{\sigma\in S}D_{B}\left(\rho,\sigma\right).\label{eq:E}\end{equation}
Here $S$ is the set of separable mixed states. With fidelity $F\left(\rho,\sigma\right)=\left(Tr\left[\sqrt{\sqrt{\rho}\sigma\sqrt{\rho}}\right]\right)^{2}$
Bures distance is defined as \begin{eqnarray}
D_{B}\left(\rho,\sigma\right) & = & 2-2\sqrt{F\left(\rho,\sigma\right)}.\label{eq:D}\end{eqnarray}
It is trivial to see that $E_{B}\left(\rho\right)$ is zero on separable
states only. Further $E_{B}$ is invariant under local unitary operations
and nonincreasing under LOCC operations, proofs can be found in \cite{Vedral1998}. 

In this paper we will give an expression for $E_{B}$ for an arbitrary
state of two qubits. This result is heavily based on results in \cite{Wootters1998}.
There the concurrence $C\left(\rho\right)$ for a two qubit state
$\rho$ was defined as \[
C\left(\rho\right)=\max\left\{ 0,\lambda_{1}-\lambda_{2}-\lambda_{3}-\lambda_{4}\right\} ,\]
where $\lambda_{i}$ are squareroots of the eigenvalues of $\rho\left(\sigma_{y}\otimes\sigma_{y}\right)\rho^{\star}\left(\sigma_{y}\otimes\sigma_{y}\right)$
in decreasing order. Complex conjugation is taken in the standard
basis, and $\sigma_{y}=\left(\begin{array}{cc}
0 & -i\\
i & 0\end{array}\right)$.
\begin{thm*}
For an arbitrary two qubit state $\rho$ holds \begin{equation}
E_{B}\left(\rho\right)=2-2\sqrt{\frac{1+\sqrt{1-C\left(\rho\right)^{2}}}{2}}.\label{eq:E-1}\end{equation}
\end{thm*}
\begin{proof}
Let $\sigma$ be a separable state that maximizes the fidelity among
all separable states. We will show that $F\left(\rho,\sigma\right)=\frac{1+\sqrt{1-C\left(\rho\right)^{2}}}{2}$
which will end the proof.

According to \cite[Theorem 9.4]{Nielsen2000} holds: \begin{equation}
F\left(\rho,\sigma\right)=\max_{\phi}\left|\braket{\psi|\phi}\right|^{2},\label{eq:F-1}\end{equation}
where $\ket{\psi}$ is a purification of $\rho$ and maximization
is done over all purifications of $\sigma$ denoted by $\ket{\phi}$.
In the following $\ket{\phi}$ will denote a particular purification
of $\sigma$ that reaches the maximum, that is $F\left(\rho,\sigma\right)=\left|\braket{\psi|\phi}\right|^{2}$.

If $\rho=\sum_{i}p_{i}\ket{\psi_{i}}\bra{\psi_{i}}$ and $\sigma=\sum_{i}q_{i}\ket{\phi_{i}^{\left(1\right)}}\bra{\phi_{i}^{\left(1\right)}}\otimes\ket{\phi_{i}^{\left(2\right)}}\bra{\phi_{i}^{\left(2\right)}}$,
then \begin{eqnarray}
\ket{\psi} & = & \sum_{i}\sqrt{p_{i}}\ket{\psi_{i}^{\left(0\right)}}\ket{\psi_{i}},\label{eq:psi}\\
\ket{\phi} & = & \sum_{i}\sqrt{q_{i}}\ket{\phi_{i}}.\label{eq:phi}\end{eqnarray}
Here $p_{i}\geq0$, $q_{i}\geq0$, $\sum_{i}p_{i}=\sum_{i}q_{i}=1$,
$\braket{\psi_{i}^{\left(0\right)}|\psi_{j}^{\left(0\right)}}=\braket{\phi_{i}^{\left(0\right)}|\phi_{j}^{\left(0\right)}}=\delta_{ij}$,
and $Tr_{0}\left[\ket{\psi}\bra{\psi}\right]=\rho$, $Tr_{0}\left[\ket{\phi}\bra{\phi}\right]=\sigma$,
$\ket{\phi_{i}}=\ket{\phi_{i}^{\left(0\right)}}\ket{\phi_{i}^{\left(1\right)}}\ket{\phi_{i}^{\left(2\right)}}$.

The states $\ket{\phi_{i}^{\left(1\right)}}$ and $\ket{\phi_{i}^{\left(2\right)}}$
can always be chosen such that $\braket{\psi|\phi_{i}}\geq0$, thus
using \prettyref{eq:phi} \begin{equation}
\left|\braket{\psi|\phi}\right|=\sum_{i}\sqrt{q_{i}}\left|\braket{\psi|\phi_{i}}\right|.\label{eq:ph}\end{equation}
Using lagrange multipliers it can easily be shown that \prettyref{eq:ph}
is maximal iff $\sqrt{q_{i}}=\frac{\left|\braket{\psi|\phi_{i}}\right|}{\sqrt{\sum_{i}\left|\braket{\psi|\phi_{i}}\right|^{2}}}$.
Using this we get \begin{equation}
\left|\braket{\psi|\phi}\right|^{2}=\sum_{i}\left|\braket{\psi|\phi_{i}}\right|^{2}.\label{eq:ph-3}\end{equation}
Note that there always is a unitary matrix $u$ such that \begin{equation}
\ket{\psi_{i}^{\left(0\right)}}=\sum_{j}u_{ij}\ket{\phi_{j}^{\left(0\right)}}.\label{eq:u}\end{equation}
Using \prettyref{eq:u} in \prettyref{eq:psi} we get \begin{eqnarray}
\ket{\psi} & = & \sum_{i}\sum_{j}\sqrt{p_{i}}u_{ij}\ket{\phi_{j}^{\left(0\right)}}\ket{\psi_{i}}\label{eq:ps}\\
 & = & \sum_{j}\sqrt{p'_{j}}\ket{\phi_{j}^{\left(0\right)}}\ket{\psi'_{j}},\nonumber \end{eqnarray}
where $\rho=\sum_{j}p'_{j}\ket{\psi'_{j}}\bra{\psi'_{j}}$. With this
result we see that it is sufficient to restrict ourselves to the case
where $\ket{\psi_{i}^{0}}=\ket{\phi_{i}^{\left(0\right)}}$. Then
\begin{equation}
\left|\braket{\psi|\phi_{i}}\right|^{2}=p_{i}\left|\braket{\psi_{i}|\phi_{i}^{\left(1\right)}}\ket{\phi_{i}^{\left(2\right)}}\right|^{2}.\label{eq:ph-1}\end{equation}
Let now \begin{equation}
\ket{\psi_{i}}=\sqrt{\lambda_{i}}\ket{u_{i}^{\left(1\right)}}\ket{u_{i}^{\left(2\right)}}+\sqrt{1-\lambda_{i}}\ket{v_{i}^{\left(1\right)}}\ket{v_{i}^{\left(2\right)}}\label{eq:schmidt}\end{equation}
be the Schmidt decomposition of $\ket{\psi_{i}}$, $\lambda_{i}\geq\frac{1}{2}$.
It is very easy to show that $\left|\braket{\psi_{i}|\phi_{i}^{\left(1\right)}}\ket{\phi_{i}^{\left(2\right)}}\right|^{2}$
is maximal for $\ket{\phi_{i}^{\left(1\right)}}=\ket{u_{i}^{\left(1\right)}}$,
$\ket{\phi_{i}^{\left(2\right)}}=\ket{u_{i}^{\left(2\right)}}$ with
the maximal value $\lambda_{i}$. Noting this we get \begin{equation}
\left|\braket{\psi|\phi_{i}}\right|^{2}=p_{i}\lambda_{i}.\label{eq:ph-4}\end{equation}
Using \prettyref{eq:ph-4} in \prettyref{eq:ph-3} we get \begin{equation}
\left|\braket{\psi|\phi}\right|^{2}=\sum_{i}p_{i}\lambda_{i}.\label{eq:ph-2}\end{equation}
From \cite{Wootters1998} we know that every two qubit state $\rho$
has a decomposition into four pure states all having equal entanglement
and thus equal Schmidt coefficients, denoted here by $\mu$ and $1-\mu$.
Further this decomposition minimizes average entanglement, which means
that with $h\left(x\right)=-x\log_{2}x-\left(1-x\right)\log_{2}\left(1-x\right)$
holds: \begin{equation}
h\left(\mu\right)\leq\sum_{i}p_{i}h\left(\lambda_{i}\right)\label{eq:h}\end{equation}
for any decomposition of the given two qubit state into pure states
with Schmidt coefficients $\lambda_{i}$ and $1-\lambda_{i}$ and
probabilities $p_{i}$. Using \prettyref{eq:h} we will now show that
\prettyref{eq:ph-2} is maximal if all $\lambda_{i}$ are equal to
$\mu$.

Suppose the opposite, that is there is a decomposition into states
having Schmidt coefficients $\lambda_{i}$ and probabilities $p_{i}$
such that $\sum_{i}p_{i}\lambda_{i}>\mu$. Then holds: \begin{equation}
\sum_{i}p_{i}h\left(\lambda_{i}\right)\leq h\left(\sum_{i}p_{i}\lambda_{i}\right)<h\left(\mu\right).\label{eq:ineq}\end{equation}
The first inequality is true because $h$ is concave, the second inequality
is true because $h\left(x\right)$ decreases for $x\geq\frac{1}{2}$.
This is in contradiction to \prettyref{eq:h}.

The consequence is that \prettyref{eq:ph-2} is maximal if all $\lambda_{i}$
are equal to $\mu$. In \cite{Wootters1998} it was shown that in
this case \begin{equation}
\mu=\lambda_{i}=\frac{1+\sqrt{1-C\left(\rho\right)^{2}}}{2}.\label{eq:lambda}\end{equation}
With \prettyref{eq:ph-2} we have showed that $F\left(\rho,\sigma\right)=\left|\braket{\psi|\phi}\right|^{2}=\frac{1+\sqrt{1-C\left(\rho\right)^{2}}}{2}$.
This completes the proof.
\end{proof}
In this paper a formula for Bures measure of entanglement for two
qubit states was presented and proved. It is interesting to note that
the proof is based on the fact that a two qubit state can always be
decomposed into pure states, all having equal entanglement. This is
in general true only for two qubits, this technique will not work
for higher dimensions \cite{Horodecki2009}.

For the most measures of entanglement presented so far no explicit
expressions are known even for two qubit states \cite{Horodecki2009}.
It is very probable that the concurrence also plays an important role
in other entanglement measures, further work is needed in this direction.

\begingroup\raggedright\endgroup

\end{document}